\documentclass[conference]{IEEEtran}
\IEEEoverridecommandlockouts
\usepackage{cite}
\usepackage{amsmath,amssymb,amsfonts}
\usepackage{algorithmic}
\usepackage{graphicx}
\usepackage{textcomp}
\usepackage{xcolor}
\usepackage{multirow}
\usepackage{subfigure}
\usepackage{makecell}
\pdfoutput=1

\def\BibTeX{{\rm B\kern-.05em{\sc i\kern-.025em b}\kern-.08em
    T\kern-.1667em\lower.7ex\hbox{E}\kern-.125emX}}
\begin{document}

\title{Low-Carbon Economic Dispatch of Bulk Power Systems Using Nash Bargaining Game\\
\thanks{This work was supported by the Science and Technology Project of State Grid Corporation of China 'Research on inter-provincial spot market clearing technology considering AC grid congestion (5108-202155045A-0-0-00)'.}
}

\author{
\IEEEauthorblockN{Xuyang Li, Guangchun Ruan, Haiwang Zhong}
\IEEEauthorblockA{\textit{Department of Electrical Engineering, Tsinghua University, Beijing 100084, China} \\
lixuyang21@mails.tsinghua.edu.cn}
}

\maketitle

\begin{abstract}
Decarbonization of power systems plays a crucial role in achieving carbon neutral goals across the globe, but there exists a sharp contradiction between the emission reduction and levelized generation cost.
Therefore, it is of great importance for power system operators to take economic as well as low-carbon factors into account. 
This paper establishes a low-carbon economic dispatch model of bulk power systems based on Nash bargaining game, which derives a Nash bargaining solution making a reasonable trade-off between economic and low-carbon objectives.
Because the Nash bargaining solution satisfies Pareto effectiveness, we analyze the computational complexity of Pareto frontiers with parametric linear programming and interpret the inefficiency of the method. 
Instead, we assign a group of dynamic weights in the objective function of the proposed low-carbon economic dispatch model so as to improve the computational efficiency by decoupling time periods and avoiding the complete computation of Pareto frontiers.
In the end, we validate the proposed model and the algorithm by a realistic nationwide simulation in mainland China.
\end{abstract}


\section{Introduction}

Electricity sector is a major contributor to the devastating global warming~\cite{b3}. It is still challenging to reduce the electricity-related carbon emissions, but the consensus is that an innovative energy management scheme~\cite{b17} as well as the flexibility resource exploration~\cite{b27} are two major solutions. The focus of this paper is on the first option, and more specifically, the state-of-the-art low-carbon economic dispatch (LCED).



Mathematically, LCED is an optimization problem with three objectives of secure, economic and low-emission operation of the power system.
Traditional multi-objective optimization methods aggregate the different objective functions into a single one.
Some representatives of this kind are the weighted sum method, the $\epsilon$-constraint method and the Nash bargaining method.
The weighted sum method assigns weights to each component of the objective vector and forms a linear aggregation of the objectives. The optimal solution of the single-objective problem should be a Pareto optimal solution of the original multi-objective optimization problem~\cite{b4}.
The $\epsilon$-constraint method optimize individually a selected objective while converting the remaining objectives into constraints by keeping their values in user-specified bounds. It can find solutions associated with non-convex parts of the Pareto frontier~\cite{b5}.
The objective of the Nash bargaining method is to minimize the product of the difference between each original objective and their respective worst value. The solution is Pareto optimal and a reasonable trade-off among the original objectives. But the problem is complicated due to the complexity of computing the Pareto frontier and the non-linearity of the objective function~\cite{b6}. The non-linearity is caused by the multiplication of decision variables.

As for the two objectives of secure and economic operation, the security constrained economic dispatch (SCED) model is usually used in academic research and practical applications~\cite{b7}. 
Besides, most researchers extend the objective function or constraints of the SCED model to consider the objective of low-carbon.
One way is to directly consider carbon emissions as a constraint based on the $\epsilon$-constraint method. Another way introduces carbon emissions into the objective function through carbon tax, carbon emission right price or carbon capture price, referring to the weighted sum method.

Most of the studies considered the hourly constraints on carbon emission in order to match the temporal resolution of SCED~\cite{b8,b10}.
Since the actual carbon emission constraints are usually annual, the configuration of hourly carbon emission constraints is vague and needs further research efforts~\cite{b12}.
Some studies considered longer time scales, such as 24 hours or 168 hours (one week)~\cite{b13,b14}. But it was still quite different from the annual time scale, and the reliability of the results depended on the accurate selecting of typical days and typical weeks.
Other researchers considered annual carbon emission constraints to avoid annual constraint decomposition. However, due to the increasing number of time coupled variables, the computational cost would subsequently increase and the research scale would be limited~\cite{b15}.

Compared with the carbon emission constraints, introducing carbon emissions into the objective function is more advantageous~\cite{b16,b18}. 
On the one hand, the weight of carbon emissions have intuitive economic meanings such as carbon tax and carbon capture price. 
On the other hand, this method can decouple time periods and avoid annual constraint decomposition. The carbon emissions of different time scales can be obtained by solving single-period problems and summing up their solutions.
However, it is difficult for the weighted sum method to make an objective trade-off between multiple objectives because the weights are determined subjectively.

The Nash bargaining method regards each objective as a game player, while the equilibrium solution is the trade-off of each objective.
Reference~\cite{b20} took power purchase cost and power generation emissions as two objectives, and obtained the corresponding relationship between the two objectives by adjusting the value of $\epsilon$ (which is equivalent to computing the Pareto frontier).
Instead, reference~\cite{b21} avoided computing the Pareto frontier through constraint relaxation, made the objective function convex , and used the outer approximation algorithm to effectively solve the problem.
However, there are still temporal coupling issues in these studies. 
This may result in a relatively heavy computational burden when conducting a long-term simulation with an increasing number of variables and constraints.


In brief, this paper makes the following contributions:
\begin{itemize}
    \item This paper establishes a low-carbon economic dispatch model of bulk power systems using Nash bargaining game, and obtains the Nash bargaining solution, which is a reasonable trade-off between economic and low-carbon objectives.
    \item A heuristic algorithm based on the weighted sum method is proposed to avoid temporal coupling. The algorithm computes the complete Pareto frontier of the problem, and obtain the Nash bargaining solution after a few iterations. Finally, we validate the model and the algorithm using a realistic nationwide simulation in mainland China.
\end{itemize}

In the remainder of this paper, Section~II describes the LCED model and Nash bargaining problem; Section~III analyzes the computational complexity of the Pareto frontier and introduces the dynamic weight algorithm; case study and conclusions are given in Section~IV and Section~V.

\section{Problem Statement and Model}

\subsection{Low-Carbon Economic Dispatch Model}
We typically establish a low-carbon economic dispatch model of bulk power systems, as shown in \eqref{eq1}-\eqref{eq6}:
\begin{align}
    \min_{p^g, p^{t l(d c)}, \theta} \quad
    & \left\{\sum_{t, r, u} c_{t, r, u} p_{t, r, u}^{g} ; \sum_{t, r, u} e_{t, r, u} p_{t, r, u}^{g}\right\}\label{eq1} \\
    \text { s.t. } \quad \quad
    & \sum_{u} \boldsymbol{p}_{t, u}^{g}-Y^{\mathrm{ac}} \boldsymbol{\theta}_{t}-W^{d c} \boldsymbol{p}_{t}^{t l(d c)}=\boldsymbol{p}_{t}^{d}, \forall t \label{eq2} \\
    & \boldsymbol{p}_{t}^{t l(a c)}=\operatorname{diag}(\boldsymbol{y}) W^{\mathrm{ac}} \boldsymbol{\theta}_{t}, \forall t \label{eq3} \\
    & -\overline{\boldsymbol{p}}^{t l(a c)} \leq \boldsymbol{p}_{t}^{t l(a c)} \leq \overline{\boldsymbol{p}}^{t l(a c)}, \forall t \label{eq4} \\
    & 0 \leq \boldsymbol{p}_{t}^{t l(d c)} \leq \overline{\boldsymbol{p}}^{t l(d c)}, \forall t \label{eq5} \\
    & \underline{p}_{t, r, u}^{g} \leq p_{t, r, u}^{g} \leq \bar{p}_{t, r, u}^{g}, \forall t, r, u\label{eq6}
\end{align}
where the subscripts $t,r,u$ represent the time period, node and unit respectively; $c,e$ are the cost coefficient and carbon emission coefficient of the unit; $p^g$ represents the output of the unit, and $p^d$ represents the nodal load; $Y^{ac}$ is the admittance matrix of AC lines; $\theta$ is the phase angle of each node; $W$ represents the node-branch incident matrix, which superscripts $dc,ac$ represent DC and AC lines, respectively; $\boldsymbol{y}$ is the mutual admittance vector of AC lines; $\overline{\boldsymbol{p}}^{tl(d c)},\boldsymbol{p}^{tl}$ represent the line capacity constraint vector and the line power flow vector; the $\underline{p}^{g},\bar{p}^{g}$ represent the lower and upper bounds of the unit output, respectively.

Assuming that the power of DC lines is adjustable, the model takes unit output, DC line power and nodal phase angle as decision variables, and aims to minimize power purchase cost and power generation emissions(as \eqref{eq1}), considering security constraints such as nodal power balance, network power flow, line capacity, and unit capacity(as \eqref{eq2}-\eqref{eq6}). 
Because the power flow directions of DC lines are usually fixed, the lower bounds of their capacity constraints are 0. Instead, AC lines are more responsible for maintaining the stability of the grid, and there may be short-term power flow reversal, so we consider the AC power flow in both positive and negative directions. 

\subsection{Nash Bargaining Game}

Generally, due to the installation of emission reduction equipment or the adoption of more advanced technologies, the LCOE of power generators with lower emissions will be higher.
Therefore there is a competition between the two objectives of the preceding model (minimizing power purchase cost and power generation emissions).
Considering the two objectives as two players in the bargaining game, they both want to optimize their payoffs, and they will eventually reach a mutually acceptable solution, which is a trade-off solution of the multi-objective optimization problem. 
Then, the original multi-objective optimization problem can be converted into a Nash bargaining problem.

In theory~\cite{b22}, the equilibrium solution of the bargaining problem should satisfy four properties:

\subsubsection{Pareto Effectiveness}
The equilibrium solution is not dominated by any other solution, which means that there is no other feasible solution, so that each target value is not inferior to the equilibrium solution, and that at least one target value is better than the equilibrium solution.

\subsubsection{Symmetry}
The equilibrium solution isn't affected by the order of the players' action, but only by their utility function. If the utility functions of the players are the same, then their utilities are the same when the equilibrium is reached.

\subsubsection{Affine Transformation Independence}
Performing affine transformation on the utility function of any player does not change the equilibrium and the equilibrium utility of other players, only resulting in the same affine transformation on the utility of the player.

\subsubsection{Independence of Irrelevant Alternatives}
Eliminating non-equilibrium strategies from the strategy space does not affect the equilibrium of the game.

The equilibrium solution satisfying the above properties is defined as \eqref{eq7}, where $d_1,d_2$ are the utilities of the players, corresponding to the objective values of the multi-objective problem; $f_1 (x),f_2 (x)$ are respectively the worst utilities of them, corresponding to the worst objective values of the multi-objective problem.
\begin{equation}
    x^{*}=\underset{x \in X}{\operatorname{argmax}}\ \left(d_{1}-f_{1}(x)\right)\left(d_{2}-f_{2}(x)\right)\label{eq7}
\end{equation}

Due to the Pareto effectiveness, we can compute the Pareto frontier to narrow the feasible region of the Nash bargaining problem.
The LCED model (as \eqref{eq1}-\eqref{eq6}) can be abbreviated as \eqref{eq8}. And computing its Pareto frontier is equivalent to solving \eqref{eq9} where $\lambda \in [0,1]$:
\begin{equation}
    \begin{aligned}
    \min\quad & \left\{c^{\text{T}} x ; e^{\text{T}} x\right\} \\
    \text { s.t. } \quad & G x \leq g, H x=h
    \end{aligned}
    \label{eq8}
\end{equation}
\begin{equation}
    \begin{aligned}
    \min\quad & \lambda c^{\text{T}} x+(1-\lambda) e^{\text{T}} x \\
    \text { s.t. } \quad & G x \leq g, H x=h        
    \end{aligned}
    \label{eq9}
\end{equation}

Let the optimal solution of \eqref{eq9} be $x^{*}(\lambda)$, then the Pareto frontier of \eqref{eq8} is $\cup_{\lambda \in[0,1]} x^{*}(\lambda)$.

\section{Methodology}

This section explains the computational complexity of computing Pareto frontiers by parametric linear programming. Instead, a dynamic weight algorithm is designed as an alternative option by scanning $\lambda$ and solve \eqref{eq9}.


\subsection{Computational Complexity Analysis of Pareto Frontier}

Equation \eqref{eq9} is a linear programming problem parameterized by $\lambda$. In theory, the parameter space can be divided into several critical regions, in which the optimal solution $x^{*}(\lambda)$ and the optimal value are linear mappings about the parameters $\lambda$~\cite{b23}.
Then the Pareto frontier $\cup_{\lambda \in[0,1]} x^{*}(\lambda)$ is a piecewise linear mapping of the parameter. In small-scale cases,
the Nash bargaining problem (as \eqref{eq7}) can be transformed into the extreme value problem of a segmented univariate quadratic function, which is more tractable.

Without loss of generality, equation \eqref{eq9} can be written in a compact form as \eqref{eq10}:
\begin{equation}
\begin{aligned}
    \min\quad &(c+E \theta)^{\text{T}} x \\
    \text { s.t. } \quad
    & A x=b \\
    & x \geq 0
    \label{eq10}
\end{aligned}
\end{equation}

Let us consider its dual problem \eqref{eq11}, which is a linear programming problem with $\theta$ on the right-hand side.
\begin{equation}
    \begin{aligned}
    \max\quad & b^{\text{T}} w \\
    \text { s.t. } \quad & A^{\text{T}} w \leq c+E \theta
    \end{aligned}
    \label{eq11}
\end{equation}

As for the dual problem \eqref{eq11}, reference~\cite{b23} has proved that if the given parameter $\theta_0$ satisfies:
\begin{equation}
    \tilde{A}^{\text{T}} w^{*}\left(\theta_{0}\right)=\tilde{c}+\tilde{E} \theta\label{eq12}
\end{equation}
\begin{equation}
    {\bar{A} ^{\text{T}}} w^{*}\left(\theta_{0}\right)<\bar{c}+\bar{E} \theta\label{eq13}
\end{equation}
Then we obtain the critical region $\Theta_{0}$, which is $CR(\theta_{0})$. And $\forall \theta \in \Theta_{0}$, the corresponding optimal solution $w^{*}(\theta)$ and optimal value $z^{*}(\theta)$ can be calculated as \eqref{eq15} and \eqref{eq16}.
\begin{equation}
    \Theta_{0}=\left\{\theta \mid\left(\bar{A}^{\text{T}}({\tilde{A}^{\text{T}}})^{-1} \tilde{E}-\bar{E}\right) \theta<\bar{c}-\bar{A}^{\text{T}} ({\tilde{A}^{\text{T}} })^{-1} \tilde{c}\right\} \label{eq14}
\end{equation}
\begin{equation}
    w^{*}(\theta)=({\tilde{A}^{\text{T}} })^{-1}(\tilde{c}+\tilde{E} \theta)\label{eq15}
\end{equation}
\begin{equation}
    z^{*}(\theta)=b^{\text{T}} w^{*}(\theta)=b^{\text{T}}({\tilde{A}^{\text{T}} })^{-1}(\tilde{c}+\tilde{E} \theta)\label{eq16}
\end{equation}

Let the set of active constraint subscripts in \eqref{eq12} be $Q$ and consider the original problem \eqref{eq10}.
By the complementary relaxation theorem, if $j \notin Q$, then $x_j=0$.
\begin{equation}
\begin{aligned}
    \min\quad & 1^{\text{T}} y\\
    \text { s.t. }\quad 
    & \sum_{j \in Q} P_{j} x_{j}+y=b\\
    & y \geq 0
\end{aligned}
    \label{eq17}
\end{equation}

Therefore, as for the problem \eqref{eq17}, there should be $y^*=0,x_j^*=a_j (j \in Q)$.
Then the optimal solution of the original problem \eqref{eq10} should be
\begin{equation}
    \left\{
    \begin{aligned}
    x_{j}^{*}& =a_{j}, \quad& \text{if } j \in Q \\
    x_{j}^{*}& =0, \quad& \text{if } j \notin Q
    \end{aligned}
    \right.
    \label{eq18}
\end{equation}

In short, for a linear programming problem with parameters in the objective function coefficients (such as \eqref{eq10}), if the given parameters $\theta_0$ satisfy \eqref{eq12} and \eqref{eq13}, the critical region including $\theta_0$, the optimal solution and the optimal value corresponding to the critical region are respectively defined by \eqref{eq14} \eqref{eq18} \eqref{eq16}, which means that the value of parameters in the same critical region leads to the same optimal solution. 
It can be seen that if the Pareto frontier of a multi-objective problem is solved by the parametric programming method (such as \eqref{eq9}), a critical region obtained only corresponds to a point on the Pareto frontier, and the computational cost of solving the critical region will be larger than solving a linear programming problem with given parameters. Therefore, the linear programming method is not advantageous over the method of traversing the parameters on this problem.

\subsection{Dynamic Weight Algorithm}

\begin{figure}
    \centering
    \centerline{\includegraphics[width=1\linewidth]{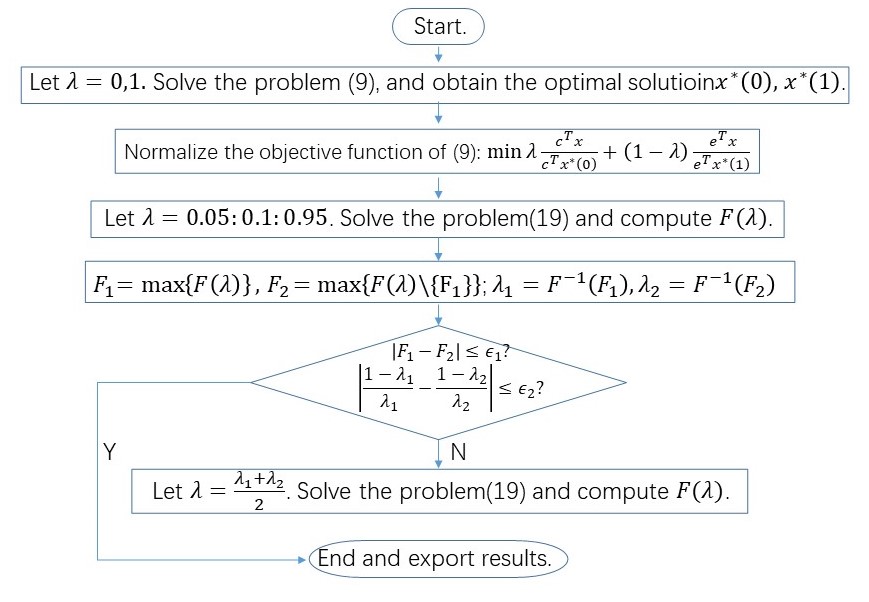}}
    \caption{Flowchart of the proposed dynamic weight algorithm.}
    \label{fig1}
\end{figure}

To derive the Pareto optimal solution with given parameters, we propose a heuristic algorithm to solve the Nash negotiation problem, which can obtain the negotiation solution while avoiding the complete Pareto frontier characterization. As shown in Fig.~\ref{fig1}, the flowchart of the proposed algorithm can be illustrated as follows:

\subsubsection{Step 1}
Let $\lambda=0,1$, solve the optimization problem \eqref{eq9}, the optimal solution is $x^* (0),x^* (1)$, then $c^{\text{T}} x^* (0),e^{\text{T}} x^* (1)$ are the worst value of the two objectives.

\subsubsection{Step 2}
Normalize the objective function of \eqref{eq9} with the worst value of the objectives, and the problem is transformed into \eqref{eq19}. 
The purpose of the transformation is to obtain a relatively uniform Pareto frontier for uniform values. Due to affine transformation independence, the normalization process does not change the Nash bargaining solution.
\begin{equation}
    \begin{aligned}
    \min\quad&  \lambda \frac{c^{\text{T}} x}{c^{\text{T}} x^{*}(0)}+(1-\lambda) \frac{e^{\text{T}} x}{e^{\text{T}} x^{*}(1)} \\
    \text { s.t. } \quad 
    & G x \leq g, H x=h
    \end{aligned}
    \label{eq19}
\end{equation}

\subsubsection{Step 3}
Let $\lambda = 0.05:0.1:0.95$, solve \eqref{eq19} and compute the objective value $F(\lambda)$(as \eqref{eq20}) of the Nash bargaining problem.
\begin{equation}
    F(\lambda)=\left(x^{*}(0)-c^{\text{T}} x^{*}(\lambda)\right)\left(x^{*}(1)-e^{\text{T}} x^{*}(\lambda)\right)\label{eq20}
\end{equation}

\subsubsection{Step 4}
The largest and second largest values of the $F(\lambda)$ sequence are marked as $F_1,F_2$, corresponding to $\lambda_1,\lambda_2$.

\subsubsection{Step 5}
Check whether the current result satisfies the convergence criterion. 
If it converges, output the solution result $F_1,\lambda_1$. 
Otherwise, let $\lambda=\frac{\lambda_1+\lambda_2}{2}$, solve the optimization problem \eqref{eq19} and the target value of the Nash bargaining problem $F(\lambda)$, and return to the fourth step (There are two piece of convergence criterion, one is that the objective function value $F(\lambda)$ converges , and the other is that the target weight ratio $\frac{1-\lambda}{\lambda}$ converges).

\section{Case Study}

\subsection{Basic Setup}
The network topology in mainland China is simplified to 31 nodes according to provincial administrative units and provincial power grid companies. 
With necessary simplifications, each node considers six equivalent units, namely thermal, gas, hydro, nuclear, wind, and photovoltaics units. The available output curves of wind and photovoltaics are processed according to the meteorological data from NREL~\cite{b25} and the annual available output of wind and photovoltaics in each province.

Parameters of DC lines and AC lines of 500kV and above are investigated.
The load curve of each node is obtained from typical load curves of each province.

\subsection{Simulation Results}

We select a week(168h) as example. We set $\epsilon_1=10^{-4}*F_{1}$ and $\epsilon_2=0.02$ to ensure convergence. The former limits the relative deviation between $F_{1}$ and $F_{2}$ so as not to consider their order of magnitudes. And the latter essentially limits the absolute deviation of the carbon tax, formulated as $\frac{1-\lambda}{\lambda}$.

\begin{figure}
	\centering
	\subfigure[Base Points] {\includegraphics[width=.24\textwidth]{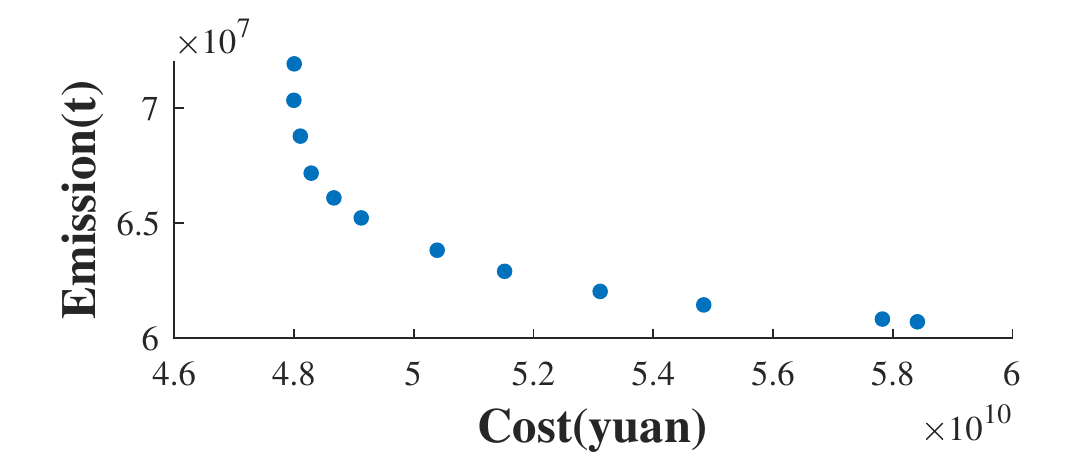}}
	\subfigure[Iteration 1] {\includegraphics[width=.24\textwidth]{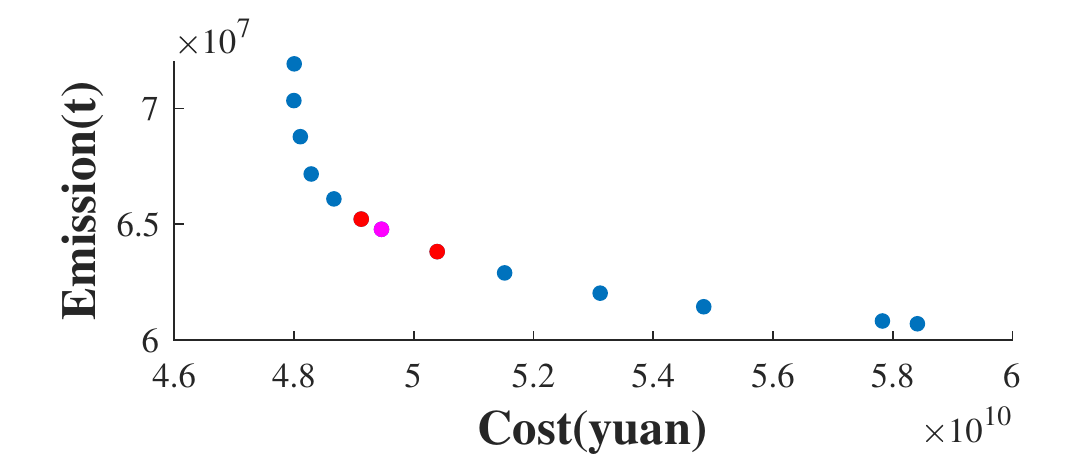}}
	\subfigure[Iteration 2] {\includegraphics[width=.24\textwidth]{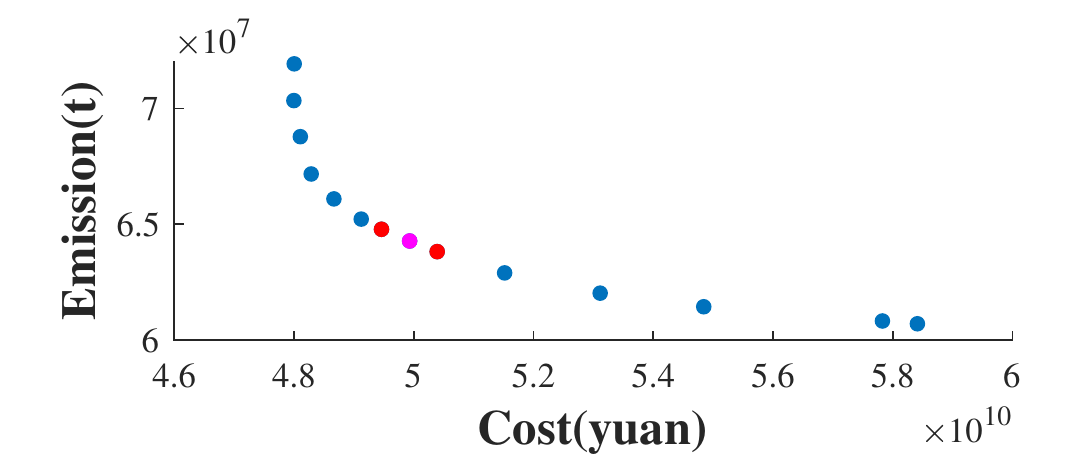}}
	\subfigure[Iteration 3] {\includegraphics[width=.24\textwidth]{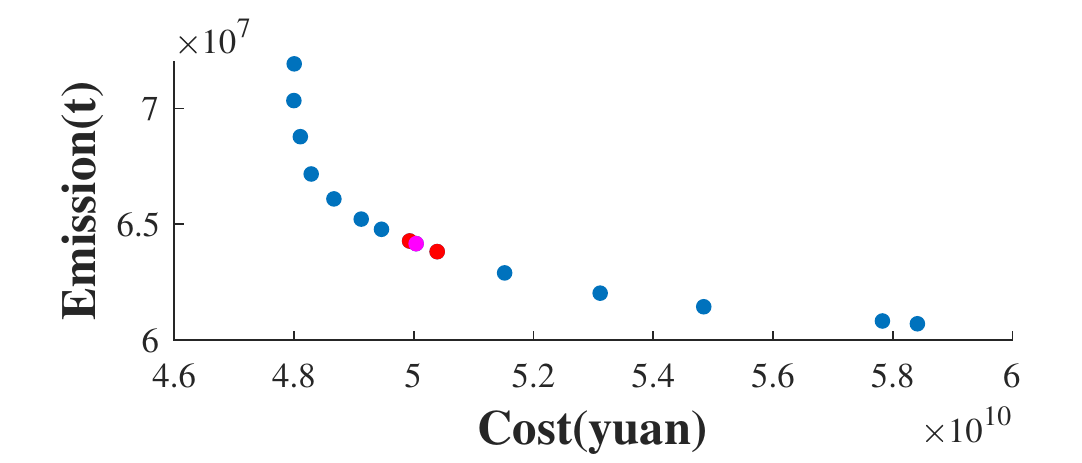}}
	\subfigure[Iteration 4] {\includegraphics[width=.24\textwidth]{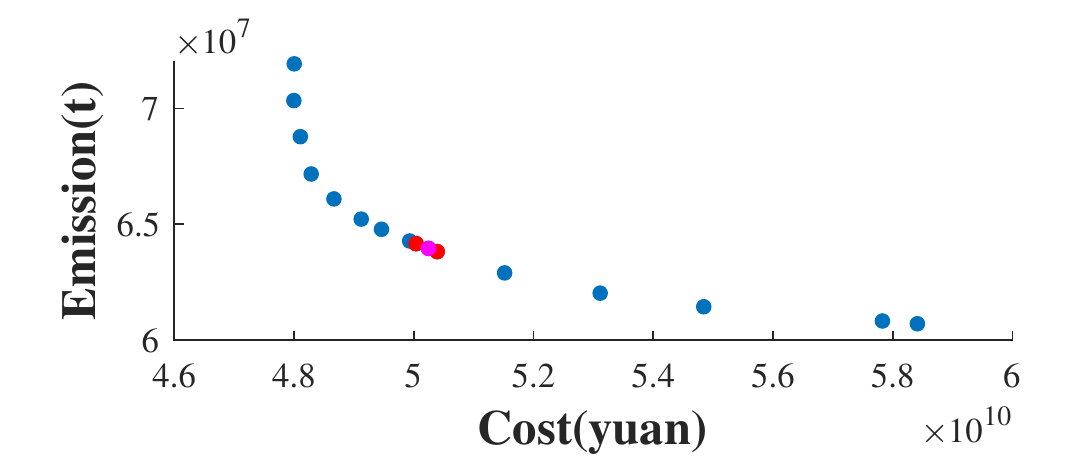}}
	\subfigure[Iteration 5] {\includegraphics[width=.24\textwidth]{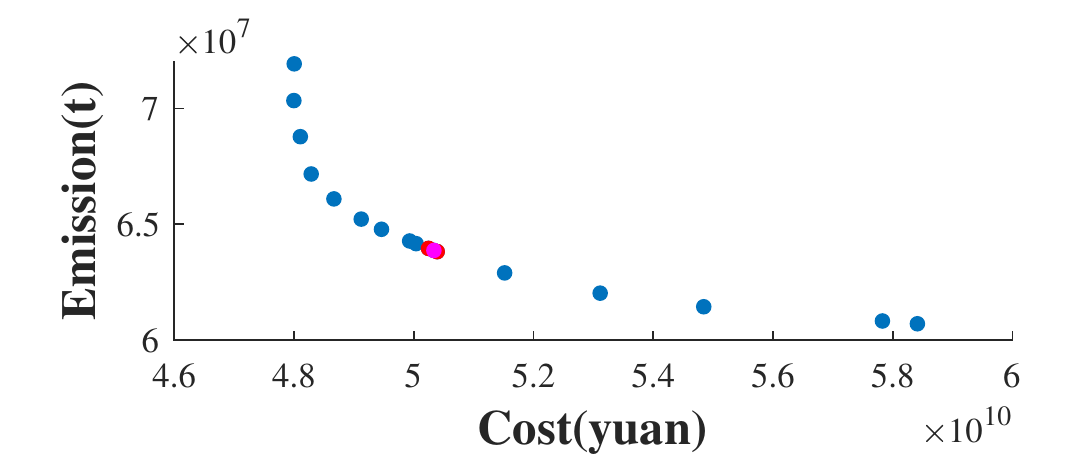}}
	\subfigure[Final Solution] {\includegraphics[width=.24\textwidth]{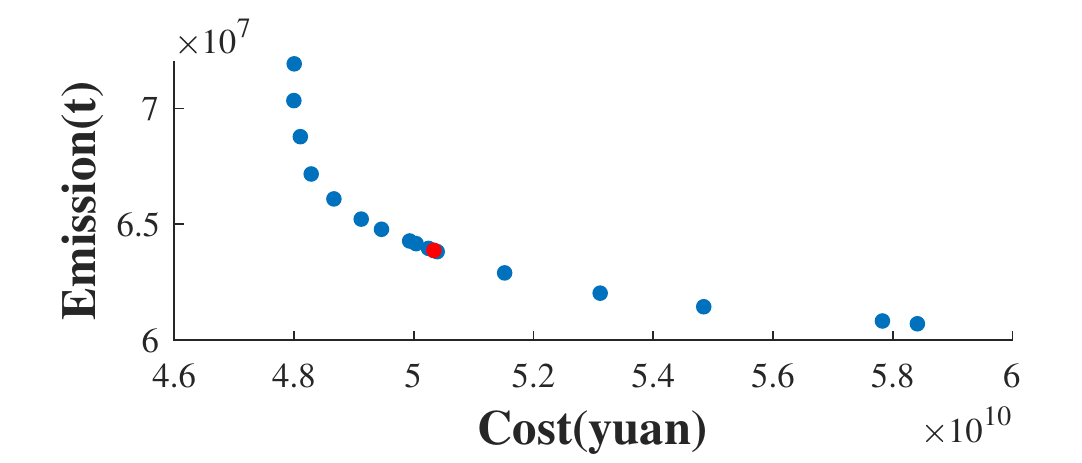}}
	\caption{Selected iteration details for the Nash Bargaining Solution.}
	\label{fig2}
\end{figure}

The computing details of the Nash bargaining solution is shown in Fig.~\ref{fig2}. To reach the convergence criterion, 5 iterations are executed. More than points calculated in iteration, there are 12 base points. However, there is a risk of missing the optimal solution in reducing base points. Besides, these points briefly delineate the Pareto frontier, which visualizes the contradiction between the emission reduction and power purchase cost.

Optimal results of different values of $\lambda$ in the iteration process are listed in Table~\ref{tab1}. With different values of $\lambda$, the power purchase cost may vary from 48.0 to 58.4 billion yuan, and the emission level may range from 60.7 to 72.0 million tons. When $\lambda=0.45625$, the Nash bargaining game reaches equilibrium. The cost is 50.2 billion yuan and the emission level is 64.0 million tons. In contrast to the case of $\lambda=0$ and $\lambda=1$, the cost and the emission level are reduced by 8.2 billion yuan and 8.0 million tons, respectively dropping about 14 \% and 11 \%.  

Overall outputs of 6 kinds of units of different values of $\lambda$ are listed in Table~\ref{tab2}. 
As the weight of the economic objective increases, the output of thermal units increases, while others decrease. Different costs may yield a great impact on the result. 
The national average feed-in tariffs of thermal, gas, hydro, nuclear, wind, and photovoltaics units are 370.52, 584.10, 267.19, 395.02, 529.01, 859.79 yuan/MWh, respectively. 
There is a small drop in the overall output of hydro units with lowest national average feed-in tariff, because feed-in tariffs of hydro units may be higher than thermal units in some provinces like Zhejiang Province.

With the Nash bargaining solution, there is $\lambda=0.45625$, and then the carbon tax is 944 yuan/ton. The carbon tax by simulated is much higher than the current price in the carbon market, equaling 54.2 yuan/ton. Thus, it can be inferred that current carbon management in mainland China is insufficient and should be reinforced.

\begin{table}
\caption{Optimal Results of Different Values of $\lambda$ in the Iteration Process}
\begin{center}
\begin{tabular}{ccccc}
\hline
Process                     & $\lambda$        & \makecell[c]{Cost\\($*10^{10}$ yuan)} & \makecell[c]{Emissions\\($*10^{7}$ ton)} & $F(*10^{16})$ \\
\hline
\multirow{2}{*}{Base Points 1}  & 0        & 5.8410    & 6.0698    & 0                          \\
                       & 1        & 4.8002    & 7.1923    & 0                          \\
\hline
\multirow{10}{*}{Base Points 2} & 0.05     & 5.7826    & 6.0818    & 0.6486                     \\
                       & 0.15     & 5.4842    & 6.1433    & 3.7433                     \\
                       & 0.25     & 5.3112    & 6.2021    & 5.2464                     \\
                       & 0.35     & 5.1516    & 6.2895    & 6.2237                     \\
                       & 0.45     & 5.0390    & 6.3813    & 6.5039                     \\
                       & 0.55     & 4.9121    & 6.5218    & 6.2274                     \\
                       & 0.65     & 4.8665    & 6.6091    & 5.6835                     \\
                       & 0.75     & 4.8286    & 6.7167    & 4.8152                     \\
                       & 0.85     & 4.8105    & 6.8778    & 3.2412                     \\
                       & 0.95     & 4.8026    & 7.0337    & 1.6515                     \\
\hline
Iteration 1                  & 0.5      & 4.9460    & 6.4779    & 6.3939                     \\
Iteration 2                  & 0.475    & 4.9930    & 6.4273    & 6.4869                     \\
Iteration 3                  & 0.4625   & 5.0038    & 6.4158    & 6.5008                     \\
Iteration 4                  & 0.45625  & 5.0245    & 6.3954    & 6.5069                     \\
Iteration 5                  & 0.453125 & 5.0335    & 6.3865    & 6.5063          \\
\hline

\end{tabular}
\label{tab1}
\end{center}
\end{table}

\begin{table}[]
\caption{Overall Outputs of 6 Kinds of Units of Different Values of $\lambda$}
\begin{center}
\begin{tabular}{ccccccc}
\hline
\multirow{2}{*}{$\lambda$} & \multicolumn{6}{c}{\begin{tabular}[c]{@{}c@{}}Outputs of Units\\ ($*10^7 $ MWh)\end{tabular}} \\ \cline{2-7} 
                        & Thermal          & Gas             & Hydro           & Nuclear          & Wind            & PV              \\ \hline
0                       & 9.5315           & 1.4897          & 1.6863          & 0.6411           & 0.6694          & 0.3153          \\
0.45625                 & 10.9993          & 0.1162          & 1.6863          & 0.6411           & 0.6694          & 0.2211          \\
1                       & 12.4283          & 0.0172          & 1.5133          & 0.3651           & 0.0066          & 0.0030 \\
\hline
\end{tabular}
\label{tab2}
\end{center}
\end{table}

\section{Conclusion}
This paper established a low-carbon economic dispatch model of bulk power systems based on Nash bargaining game, which derived a Nash bargaining solution making a reasonable trade-off between economic and low-carbon objectives.
To improve the computational efficiency, we proposed the dynamic weight algorithm.
We further validated the model and the algorithm using a realistic nationwide case derived from the real-world power systems in mainland China.
In the case, we found that the carbon tax by simulated was much higher than the current price in the carbon market. One could infer that carbon management in mainland China should be reinforced, and the carbon price is expected to rapidly rise up because of the carbon neutral requirements.

\section*{Acknowledgment}

This work was partly supported by the National Natural Science Foundation of China (No. 52122706).
And Xuyang Li would like to thank Feng Liu for recommending reference ~\cite{b20,b21} in the course of Engineering Game Theory.

\bibliographystyle{IEEEtran}
\bibliography{mybib}

\end{document}